\documentclass[letterpaper]{JHEP3}

\usepackage{amsmath,amsthm, amssymb}

\newcommand{\dst}{\displaystyle}

\newcommand{\be}{\begin{equation}}
\newcommand{\ee}{\end{equation}}
\newcommand{\ba}{\begin{array}}
\newcommand{\ea}{\end{array}}

\newcommand{\bea}{\begin{eqnarray}}
\newcommand{\eea}{\end{eqnarray}}
\newcommand{\bma}{\begin{matrix}}
\newcommand{\ema}{\end{matrix}}
\newcommand{\nn}{\nonumber}
\newcommand{\mc}{\mathcal}
\newcommand{\beq}{\stackrel{+0}{=}}

\newcommand{\st}{\sqrt2}

\newcommand{\p}{\partial}
\newcommand{\ov}{\overline}

\newcommand{\wt}{\widetilde}
\newcommand{\Eta}{\mathcal H}
\newcommand{\psibar}{\ov \psi}
\newcommand{\etabar}{\ov \eta}
\newcommand{\sibar}{\ov \sigma}
\newcommand{\labar}{\ov \la}

\newcommand{\tabar}{\ov \ta}
\newcommand{\chibar}{\ov\chi}
\newcommand{\Qbar}{\ov Q}
\newcommand{\Dbar}{\ov D}

\newcommand{\al}{\alpha}
\newcommand{\la}{\lambda}
\newcommand{\da}{\delta}

\newcommand{\Ga}{\Gamma}
\newcommand{\La}{\Lambda}
\newcommand{\Si}{\Sigma}
\newcommand{\si}{\sigma}
\newcommand{\ta}{\theta}

\title{\bf Supersymmetric bulk-brane coupling\\ with odd gauge fields}
\author{Dmitry V. Belyaev\\
Deutsches Elektronen-Synchrotron\\
DESY-Theory, Notkestrasse 85 \\
22603 Hamburg, Germany \\
E-mail: \email{dmitry.belyaev@desy.de}}

\preprint{DESY 06-131}

\abstract{
Supersymmetric bulk-brane coupling in Horava-Witten and
Randall-Sundrum scenarios, when considered 
in the orbifold (``upstairs'') picture,
enjoys similar features:
a modified Bianchi identity and a modified supersymmetry
transformation for the ``orthogonal'' part of the gauge field.
Using a toy model with a 5D vector multiplet in the bulk
(like in Mirabelli-Peskin model, but with an \emph{odd} gauge
field~$A_m$), we explain how these features arise from 
the superfield formulation. We also show that the corresponding
construction in the boundary (``downstairs'')
picture requires introduction of
a special ``compensator'' (super)field.
}

\begin{document}
%%%%%%%%%%%%%%%%%%%%%%%%%%%%%%%%%%%%%%%%%%%%%%%%%%%%%%%%%%%%%%%

%\maketitle
%\thispagestyle{empty}
\numberwithin{equation}{section}

%\newpage
%\tableofcontents

%%%%%%%%%%%%%%%%%%%%%%%%%%%%%%%%%%%%%%%%%%%%%%%%%%%%%%%%%%%%%%%
%\newpage
\section{Introduction}
%%%%%%%%%%%%%%%%%%%%%%%%%%%%%%%%%%%%%%%%%%%%%%%%%%%%%%%%%%%%%%%

The last decade saw a revival of interest in theories with
extra dimensions and brane-world scenarios. 
In 1996, Horava and Witten \cite{hw} showed 
that 11D supergravity on a manifold with boundary 
(or on $S^1/\mathbb{Z}_2$ orbifold) arises as a low energy 
limit of the strongly coupled heterotic string theory.
Three years later,
Randall and Sundrum \cite{rs} demonstrated that a simpler
5D construction with a cosmological constant in the bulk is
sufficient to naturally solve the gauge hierarchy problem
and leads to interesting phenomenological consequences.

A minimal supersymmetric version of the Randall-Sundrum scenario,
with just the tension terms on the branes, is by now well understood
\cite{srs, bkvp, bb1, my2}. Some progress has been made in including
additional matter fields on the branes (see ref.~\cite{falk}
and references therein), but the construction is not yet complete.
One interesting observation of ref.~\cite{falk} is that the
supersymmetric bulk-brane coupling (in the orbifold picture)
in both the 11D and 5D cases has two basic features in common:
\begin{enumerate} 
\item[1)]
the field strength of the bulk gauge field is shifted so that
it satisfies a modified Bianchi identity; and
\item[2)]
supersymmetry transformation of the ``orthogonal'' component
of the gauge field ($C_{11AB}$ in 11D and $B_5$ in 5D) is modified
accordingly. 
\end{enumerate}
Besides these modifications, the simplest version
of the coupling (to the 2-Fermi order) requires only adding the
Noether coupling term to the free brane Lagrangian.

Supersymmetric bulk-brane coupling can be nicely formulated using
4D superfields. The original idea is due to Mirabelli
and Peskin \cite{mp} (who worked with supermultiplets instead
of superfields); in the superfield language, the method was developed
and generalized to dimensions higher than five in ref.~\cite{agw}.
Although this method has already been widely used, the basic
features of the bulk-brane coupling listed above have not yet been 
explained by it. In this paper, we will fill in the gap.

In our discussion, we will use the toy (globally supersymmetric)
model of Mirabelli and Peskin, with an abelian 5D vector multiplet
in the bulk. In the orbifold picture, the 4D vector ($A_m$)
and the 4D scalar ($A_5$) components of the gauge field $A_M$
have opposite parities. Instead of choosing $A_m$ to be even,
as in ref.~\cite{mp}, we will choose it to be \emph{odd}
to make contact with the supergravity constructions (where
$C_{ABC}$ in 11D and $B_m$ in 5D are odd). This also flips the parities
of the 4D superfields used to describe the 5D vector multiplet,
compared to refs.~\cite{mp, agw}. We will find that this model
reproduces the features of the supergravity bulk-brane coupling
surprisingly well.

Our key results are as follows.
In the orbifold picture (OP), recovering the right 
component structure of the coupling from its superfield form
requires a certain field redefinition
that makes all bulk fields except $A_5$ non-singular.
In the boundary picture (BP), the singularity of $A_5$ 
is replaced by the presence of a special boundary compensator $K$.
In both cases, the boundary condition on the odd part of the gauge
field is $A_m=J_m$, where $J_m$ is a function of the brane/boundary
fields. This boundary condition is required for supersymmetry
of the action in the BP, but not in the OP.

The structure of the paper is best seen from the Contents.
We note here only that the agreement with the supergravity constructions
is achieved in Section \ref{sec-sA5}, and more explicitly in
the example of Section \ref{sec-ex}.

%%%%%%%%%%%%%%%%%%%%%%%%%%%%%%%%%%%%%%%%%%%%%%%%%%%%%%%%%%%%%%%
%%%%%%%%%%%%%%%%%%%%%%%%%%%%%%%%%%%%%%%%%%%%%%%%%%%%%%%%%%%%%%%
%\newpage
\section{Mirabelli-Peskin model with odd $A_m$}
\label{sec-rev}
%%%%%%%%%%%%%%%%%%%%%%%%%%%%%%%%%%%%%%%%%%%%%%%%%%%%%%%%%%%%%%%
%%%%%%%%%%%%%%%%%%%%%%%%%%%%%%%%%%%%%%%%%%%%%%%%%%%%%%%%%%%%%%%

In this section, we review the essentials of the
5D gauge supermultiplet, both in the component formulation and 
using 4D $N=1$ superfields; OP and BP are defined here.
Our conventions are the same as in ref.~\cite{my1}; 
supersymmetry conventions follow closely those 
of Wess and Bagger \cite{wb}.

%%%%%%%%%%%%%%%%%%%%%%%%%%%%%%%%%%%%%%%%%%%%%%%%%%%%%%%%%%%%%%%
\subsection{5D vector multiplet}
%%%%%%%%%%%%%%%%%%%%%%%%%%%%%%%%%%%%%%%%%%%%%%%%%%%%%%%%%%%%%%%

The abelian 5D gauge supermultiplet consists of a gauge field
$A_M$ ($M=0,1,2,3,5$), a real scalar $\Phi$, a symplectic-Majorana
spinor (gaugino) $\La_i$ ($i=1,2$), and a triplet of real auxiliary fields 
$X_a$ ($a=1,2,3$). The Lagrangian for this multiplet is 
\bea
\label{L5}
\mc{L}_5=-\frac{1}{4}F_{MN}F^{MN}-\frac{1}{2}\p_M\Phi\p^M\Phi
-\frac{i}{2}\wt\La^i\Ga^M\p_M\La_i+\frac{1}{2}X_a X_a \; .
\eea
The corresponding supersymmetry transformations are
\bea
%\label{MPsusytr}
\da_\Eta A_M &=& i\wt\Eta^i\Ga_M\La_i \nn\\
\da_\Eta\Phi &=& i\wt\Eta^i\La_i \nn\\
\da_\Eta X_a &=& \wt\Eta^i(\si_a)_i{}^j\Ga^M\p_M\La_j \nn\\[5pt] 
\da_\Eta\La_i &=& \da^\prime_\Eta\La_i+\da^{\prime\prime}_\Eta\La_i ,
\eea
where we made the following split,
\bea
\da^\prime_\Eta\La_i=(\Si^{MN}F_{MN}+\Ga^M\p_M\Phi)\Eta_i, \quad
\da^{\prime\prime}_\Eta\la_i=X_a(\si_a)_i{}^j\Eta_j ,
\eea
separating out the auxiliary part of the transformation.
(The supersymmetry parameter $\Eta_i$ is a constant 
symplectic-Majorana spinor.)
Under the supersymmetry transformations, the Lagrangian
varies into a total derivative. 
This is to be compared with the general variation, when
one also finds a total derivative plus terms that vanish
only when equations of motion (EOM) are used.
The total derivatives in these two
cases are similar, but differ in the fermionic parts.
In the case at hand, we find
\bea
\da\mc{L}_5 = (\text{EOM})+ \p_M K^M , \quad
\da_\Eta\mc{L}_5 =
\p_M \wt K^M ,
\eea
where
\bea
K^M &=& -F^{MN}\da A_N-\da\Phi\p^M\Phi
-\frac{i}{2}\wt\La^i\Ga^M\da\La^i
\nn\\
\wt K^M &=&
-F^{MN}\da_\Eta A_N-\da_\Eta\Phi\p^M\Phi
+\frac{i}{2}\wt\La^i\Ga^M\da^\prime_\Eta\La_i
-\frac{i}{2}\wt\La^i\Ga^M\da^{\prime\prime}_\Eta\La_i .
\eea
The total derivatives are irrelevant on the orbifold,
but essential in the boundary picture.

%%%%%%%%%%%%%%%%%%%%%%%%%%%%%%%%%%%%%%%%%%%%%%%%%%%%%%%%%%%%%%%
\subsection{OP, BP and $N=1$ supersymmetry}
%%%%%%%%%%%%%%%%%%%%%%%%%%%%%%%%%%%%%%%%%%%%%%%%%%%%%%%%%%%%%%%

In the orbifold picture (OP), the 5D space is $\mathbb{R}^{1,4}$
with a $\mathbb{Z}_2$ symmetry realized as a reflection 
$x^5\equiv z \rightarrow -z$. The ``fixed point'' at $z=0$
is a 4D plane that we call a ``brane''. In the boundary picture (BP),
the 5D space is $\mc{M}=\mathbb{R}^{1,3}\times[0,+\infty)$,
with boundary at $z=0$. In both cases, it is convenient to
make the ``$5 \rightarrow 4$'' split, 
using $M=\{m,5\}$ ($m=0,1,2,3$), and to convert 
symplectic-Majorana spinors into pairs of two-component spinors:
$\La_i \rightarrow (\la_1, \la_2)$, 
$\Eta_i \rightarrow (\eta_1, \eta_2)$.
This leads to the following form of the Lagrangian,~\footnote{
In our notation, 
$F_{mn}=\p_m A_n-\p_n A_m$, 
$F_{m5}=\p_m A_5-\p_5 A_m$, 
$X_{12}=X_1+i X_2$.
}
\bea
\label{L5mn}
\mc{L}_5 &=& -\frac{1}{4}F_{mn}F^{mn}-\frac{1}{2}F_{m5}F^{m5}
-\frac{1}{2}\p_m\Phi\p^m\Phi -\frac{1}{2}\p_5\Phi\p_5\Phi
+\frac{1}{2}X_{12}X_{12}^\ast +\frac{1}{2}X_3^2 \nn\\
&&-\left[
\frac{i}{2}\la_1\si^m\p_m\labar_1
+\frac{i}{2}\la_2\si^m\p_m\labar_2
+\frac{1}{2}(\la_2\p_5\la_1-\la_1\p_5\la_2) +h.c.
\right] .
\eea
When a brane/boundary is present, we can preserve only a half of
the $N=2$ supersymmetry parametrized by $\eta_1$ and $\eta_2$.
Without loss of generality, we set
\bea
%\label{eta}
\boxed{ \rule[-5pt]{0pt}{16pt} \quad
\eta_1=\eta, \quad \eta_2=0 .
\quad}
\eea
This gives the following $N=1$ supersymmetry transformations,
\bea
\label{bulksusy}
\da_\eta A_m &=& i \eta\si_m\labar_1+h.c. \nn\\[5pt]
\da_\eta A_5 &=& -\eta\la_2+h.c. \nn\\[5pt]
\da_\eta\Phi &=& -i\eta\la_2+h.c. \nn\\[5pt]
%%%%%
\da_\eta\la_1 &=& \si^{mn}\eta F_{mn}+i(X_3-\p_5\Phi)\eta 
\nn\\[5pt]
\da_\eta\la_2 &=& -i\si^m\etabar F_{m5}
-\si^m\etabar\p_m\Phi-i X_{12}\eta
\nn\\[5pt]
%%%%%
\da_\eta X_{12} &=& 
2i\etabar \p_5\labar_1
-2\etabar\sibar^m\p_m\la_2
\nn\\[5pt]
\da_\eta X_3 &=& 
-i\eta\p_5\la_2-\eta\si^m\p_m\labar_1
+h.c. 
\eea
In the orbifold picture, we must choose parity assignments
for all fields and parameters. 
We are interested in the
case when $\eta_1$ is even and $A_m$ is odd,
which leads to the following set of assignments,~\footnote{
Under the $\mathbb{Z}_2$ reflection, each field $f(x,z)$ is mapped into
$f(x,-z)=P[f]f(x,+z)$ with $P[f]=\pm 1$. We call
$P[f]=+1$ fields ``even'' and $P[f]=-1$ fields ``odd''.
} 
\bea
\boxed{ \rule[-5pt]{0pt}{16pt} \quad
\text{even}: A_5, \; \Phi, \; \la_2,  \; X_{12}, \; \eta_1 \qquad 
\text{odd}: A_m, \; \la_1, \; X_3, \; \eta_2  \; .
\quad}
\eea
With these assignments, the Lagrangian is even, whereas the equations
of motion and supersymmetry transformations are parity covariant.

%%%%%%%%%%%%%%%%%%%%%%%%%%%%%%%%%%%%%%%%%%%%%%%%%%%%%%%%%%%%%%%
\subsection{$N=1$ superfields $\bf V_2$ and $\bf\Phi_2$}
%%%%%%%%%%%%%%%%%%%%%%%%%%%%%%%%%%%%%%%%%%%%%%%%%%%%%%%%%%%%%%%

$N=1$ supersymmetry is most conveniently described in terms of
$N=1$ superfields. For the 5D vector multiplet, we need two
4D $N=1$ superfields: a gauge superfield $\bf V_2$ and a chiral
superfield $\bf \Phi_2$.~\footnote{
The subscript ``2'' on $\bf V_2$ and $\bf \Phi_2$ indicates
that these are bulk superfields. We reserve $\bf V$ and $\bf \Phi$
for denoting brane-localized superfields.
}
If we take $\bf V_2$ in the Wess-Zumino (WZ) gauge
(see Appendix~\ref{app2} for a discussion of this choice),
then the relation between the component bulk fields
and components of the superfields is given by \cite{agw, heb}
\bea
\label{V2WZ}
&{\bf V_2} = (0, \quad 0, \quad 0; \quad 
           A_m, \quad \la_1, \quad X_3-\p_5\Phi) &
\nn\\[5pt]
& {\bf\Phi_2} = (\Phi+i A_5, \quad 
-i\st\la_2, \quad
-X_{12}) . &
\eea
Here and henceforth we represent superfields by listing their components
in a definite order (see Appendix \ref{app1}). 
The supersymmetry transformations
(\ref{bulksusy}) are reproduced by the following superfield
transformations,
\bea
\label{V2susy}
\da_\eta {\bf V_2} &=& (\eta Q+\etabar\Qbar) {\bf V_2}
+{\bf\La_2}(\eta)+{\bf\La_2}(\eta)^\dagger \nn\\[5pt]
\da_\eta {\bf \Phi_2} &=& (\eta Q+\etabar\Qbar) {\bf \Phi_2}
+2\p_5{\bf\La_2}(\eta) ,
\eea
where the compensating gauge transformation 
(keeping $\bf V_2$ in the WZ gauge)
is given by
\bea
\label{WZcomp}
{\bf\La_2}(\eta)=\left( 0, \quad
\frac{1}{\st}\si^m\etabar A_m, \quad
-i\etabar\labar_1 \right).
\eea
Similarly, the bulk $U(1)$ gauge transformation,
\bea
\label{U1gtr}
\da_u A_M=\p_M u
\quad \Leftrightarrow \quad
\da_u A_m=\p_m u, \quad 
\da_u A_5=\p_5 u ,
\eea
is reproduced by the superfield gauge transformation
\bea
\label{V2gauge}
\da_u {\bf V_2}={\bf\La_2}(u)+{\bf\La_2}(u)^\dagger, \quad
\da_u {\bf \Phi_2}=2\p_5{\bf\La_2}(u) 
\eea
with the following parameter,
\bea
{\bf\La_2}(u)=\left(\frac{i}{2}u, \; 0, \; 0 \right).
\eea

%%%%%%%%%%%%%%%%%%%%%%%%%%%%%%%%%%%%%%%%%%%%%%%%%%%%%%%%%%%%%%%
\subsection{Superfield Lagrangian}
%%%%%%%%%%%%%%%%%%%%%%%%%%%%%%%%%%%%%%%%%%%%%%%%%%%%%%%%%%%%%%%

The Lagrangian $\mc{L}_5$ is gauge invariant and, therefore, should
be constructed out of gauge invariant superfields. Two basic gauge
invariant superfields are~\footnote{
We hide the spinor index $\al$ on $\bf W_2$
by contracting it with another spinor.
For the definition of the supersymmetry operator $Q_\al$
and the covariant superspace derivative $D_\al$, see ref.~\cite{wb}.
}
\bea
{\bf Z_2}=\p_5 {\bf V_2}-\frac{1}{2}({\bf\Phi_2}+{\bf\Phi_2}^\dagger), 
\quad
\eta{\bf W_2}=-\frac{1}{4}\eta^\al\Dbar\Dbar D_\al{\bf V_2} \; .
\eea
Their components are related to the bulk fields in the following way,
\bea
\label{Z2c}
{\bf Z_2} &=& \Big( -\Phi, \quad \la_2, \quad -i X_{12}; \quad
           -F_{m5}, \quad \p_5\la_1, \quad \p_5(X_3-\p_5\Phi) \;
               \Big) 
\nn\\
\eta{\bf W_2} &=& \Big( -i\eta\la_1, \quad
  \frac{1}{\st}\big[\eta (X_3-\p_5\Phi)+i\si^{mn}\eta F_{mn}\big], \quad 
  \eta\si^m\p_m\labar_1 \; \Big) .
\eea
The superfield Lagrangian that reproduces $\mc{L}_5$ up to
a total derivative is given by
\bea
\mc{L}_5^\prime
&=&
\frac{1}{4} \int d^2\ta \; {\bf W_2}^2 +h.c. + 
\int d^2\ta d^2\tabar \; {\bf Z_2}^2 
\nn\\
&=& \int d^2\ta d^2\tabar\Big[
\frac{1}{8} {\bf V_2} D^\al\Dbar\Dbar D_\al {\bf V_2} 
+ {\bf Z_2}^2 \Big],
\eea
where the second form is particularly suited for deriving superfield
equations of motion (and boundary conditions) and allows us to omit 
the overall superspace integration.\footnote{
The superspace integral $\int d^2\ta d^2\tabar$ is implicit in the
expressions for actions and Lagrangians in the rest of the paper.
Note also that we omit total $\p_m$ derivatives, as they are
irrelevant in both the orbifold and boundary pictures.
Total $\p_5$ derivatives, however, are kept.
}
Writing $\mc{L}_5^\prime$ in components and comparing with $\mc{L}_5$,
eq.~(\ref{L5mn}), we find
\bea
%\label{Ypr}
\mc{L}_5^\prime=\mc{L}_5-\p_5 Y^\prime, \quad
Y^\prime=\Phi(X_3-\p_5\Phi)+\frac{1}{2}(\la_1\la_2+h.c.).
\eea
In terms of the actions, on a manifold with boundary
$\mc{M}=\mathbb{R}^{1,3}\times[0,+\infty)$, we have
\bea
\label{S5p}
\boxed{ \rule[-5pt]{0pt}{16pt} \quad
S_5^\prime=\int_\mc{M} \mc{L}_5^\prime, \quad
S_5=\int_\mc{M} \mc{L}_5 \quad \Rightarrow \quad
S_5^\prime=S_5+\int_{\p\mc{M}}Y^\prime .
\quad}
\eea

%%%%%%%%%%%%%%%%%%%%%%%%%%%%%%%%%%%%%%%%%%%%%%%%%%%%%%%%%%%%%%%
%%%%%%%%%%%%%%%%%%%%%%%%%%%%%%%%%%%%%%%%%%%%%%%%%%%%%%%%%%%%%%%
\section{Bulk-brane coupling in superfields}
\label{sec-sfcoup}
%%%%%%%%%%%%%%%%%%%%%%%%%%%%%%%%%%%%%%%%%%%%%%%%%%%%%%%%%%%%%%%
%%%%%%%%%%%%%%%%%%%%%%%%%%%%%%%%%%%%%%%%%%%%%%%%%%%%%%%%%%%%%%%

In this section, we construct supersymmetric coupling 
of the bulk 5D gauge multiplet to the brane/boundary. 
The coupling gives rise to a boundary condition 
${\bf V_2}\beq {\bf J}$, 
where $\bf J$ is a function of brane localized
superfields $\bf V$, $\bf\Phi$, and a special compensator superfield 
$\bf K$ that, in the orbifold picture, corresponds to the
singular part of $\bf \Phi_2$.

%%%%%%%%%%%%%%%%%%%%%%%%%%%%%%%%%%%%%%%%%%%%%%%%%%%%%%%%%%%%%%%
\subsection{Boundary picture}
%%%%%%%%%%%%%%%%%%%%%%%%%%%%%%%%%%%%%%%%%%%%%%%%%%%%%%%%%%%%%%%

In the previous section, we arrived at the following superfield
action on a manifold with boundary $\mc{M}$,
\bea
%\label{S5pr}
S_5^\prime=\int_\mc{M} \Big[
\frac{1}{8} {\bf V_2} D\Dbar^2 D {\bf V_2} 
+ {\bf Z_2}^2 \Big] .
\eea
Its general variation gives
\bea
\da S_5^\prime=\int_\mc{M} \Big[
-{\bf Z_2}\da{\bf \Phi_2}+h.c.+\big(
\frac{1}{4}D\Dbar^2 D {\bf V_2}-2\p_5 {\bf Z_2} \big)\da {\bf V_2} \Big]
-\int_{\p\mc{M}} 2 {\bf Z_2}\da {\bf V_2}.
\eea
The bulk equations of motion, therefore, are~\footnote{
Equations of motion for chiral superfields are found by
applying $\Dbar\Dbar$ to what comes out from the general variation.
See ref.~\cite{wb} for more details.
} 
\bea
\label{P2eom}
\Dbar\Dbar {\bf Z_2}=0, \quad 
\frac{1}{4}D\Dbar^2 D {\bf V_2}-2\p_5 {\bf Z_2} =0,
\eea
while the natural boundary condition, obtained by requiring the
boundary piece of $\da S_5^\prime$ to vanish for arbitrary $\da {\bf V_2}$, 
is~\footnote{
The symbol $\beq$ is used to denote boundary conditions in
both the boundary and orbifold pictures. In the orbifold picture,
it means ``on the positive side of the brane'', at $z=+0$.
}
\bea
{\bf Z_2} \beq 0.
\eea
This is the reason why $S_5^\prime$ is the right action for lifting
on the orbifold with odd ${\bf Z_2}$ 
(that is, with even ${\bf V_2}$ and odd ${\bf \Phi_2}$).
Coupling bulk fields to
brane localized matter would make this boundary condition 
inhomogeneous \cite{my1}.

In this paper, we are interested in the other case, when
${\bf V_2}$ is odd and ${\bf \Phi_2}$ is even 
(therefore, ${\bf Z_2}$ is even).
The appropriate action is easy to guess. We define
\bea
\label{S5pp}
\boxed{ \rule[-5pt]{0pt}{16pt} \quad
S_5^{\prime\prime}=S_5^\prime+\int_{\p\mc{M}} 2 {\bf Z_2 V_2} .
\quad}
\eea
Its general variation gives the same equations of motion in
the bulk, but the boundary term and, therefore, the natural
boundary condition are now different:
\bea
\da S_5^{\prime\prime}= (\text{EOM})
+\int_{\p\mc{M}} 2 {\bf V_2}\da {\bf Z_2} 
\quad \Rightarrow \quad
{\bf V_2} \beq 0 .
\eea
This shows that $S_5^{\prime\prime}$ is the right action for lifting
on the orbifold with odd $\bf V_2$.

Adding boundary interaction that leads to the boundary condition
${\bf V_2}\beq {\bf J}$ is now straightforward. For the complete
bulk-plus-boundary action we take
\bea
\label{SJ}
S &=&
S_5^{\prime\prime}
+\frac{1}{2}\int_{\p\mc{M}}\mc{L}_4
-\int_{\p\mc{M}} 2 {\bf Z_2 J} \nn\\
&=&
S_5^\prime
+\frac{1}{2}\int_{\p\mc{M}}\mc{L}_4
+\int_{\p\mc{M}} 2 {\bf Z_2 (V_2-J)} ,
\eea
where $\mc{L}_4$ is a part of the boundary Lagrangian
that does not depend on the bulk fields and is supersymmetric
on its own. The general variation of the action gives the
required boundary condition:
\bea
\da S= (\text{EOM})
+\int_{\p\mc{M}} 2 ({\bf V_2-J})\da {\bf Z_2} 
\quad \Rightarrow \quad
\boxed{ \rule[-5pt]{0pt}{16pt} \quad
{\bf V_2} \beq {\bf J} .
\quad}
\eea
However, despite being written in terms of superfields,
the action is not yet guaranteed to be supersymmetric.
Supersymmetry transformations of $\bf V_2$ and $\bf \Phi_2$,
given in eq.~(\ref{V2susy}), are a combination of the
standard piece (with the linear supersymmetry operator acting
on them) and a special gauge transformation. As a result,
the action can be supersymmetric only when it is gauge
invariant. With $\bf V_2$ appearing in the action explicitly, 
this can be achieved only if the gauge and supersymmetry transformations
of $\bf J$ match those of $\bf V_2$. That is, the action is
supersymmetric provided $\bf J$ transforms as follows,
\bea
\label{susyJ}
\da_u {\bf J}=
{\bf \La_2^{(+)}}(u)+{\bf \La_2^{(+)}}(u)^\dagger, \quad
\da_\eta {\bf J}=(\eta Q+\etabar\Qbar){\bf J}
+{\bf \La_2^{(+)}}(\eta)+{\bf \La_2^{(+)}}(\eta)^\dagger ,
\eea
where the superscript ``(+)'' indicates restriction of the bulk quantity to
the boundary.

One can construct $\bf J=J(V,\Phi)$ with the above
transformation laws. However, this inevitably requires relating
bulk and boundary gauge invariances and leads to a rather strange
form of the coupling. Another way to satisfy eq.~(\ref{susyJ}),
motivated by the orbifold picture construction (see below), is to
introduce a special boundary superfield $\bf K$ with the following
transformation properties,
\bea
\label{susyK}
\da_u {\bf K}={\bf\La_2^{(+)}}(u), \quad
\da_\eta {\bf K}=(\eta Q+\etabar\Qbar){\bf K}
+{\bf \La_2^{(+)}}(\eta) .
\eea
If we now define
\bea
\label{JKG}
\boxed{ \rule[-5pt]{0pt}{18pt} \quad
{\bf J}={\bf K +K^\dagger +G},
\quad}
\eea
with $\bf G=G(V,\Phi)$ transforming as a gauge invariant quantity,
\bea
\label{susyG}
\da_u {\bf G}= 0, \quad
\da_\eta {\bf G}=(\eta Q+\etabar\Qbar){\bf G} ,
\eea
then $\bf J$ transforms precisely as in eq.~(\ref{susyJ}).
This way we do not need to relate the bulk gauge transformation
to a boundary one, which means that introducing the superfield $\bf K$
increases gauge symmetry of the action. Therefore, we can call $\bf K$
a ``compensator'' superfield.  

With the superfield $\bf K$ present, we do not need a boundary gauge 
transformation, so that, for example, $\bf G=\Phi^\dagger \Phi$
is a valid choice. Note also that $\bf K$ does not appear in $\mc{L}_4$,
but comes only with $\bf J$. As a result, its equation of motion is
\bea
\Dbar\Dbar {\bf Z_2} \beq 0 .
\eea
As this coincides with the restriction of the bulk equation of motion 
for $\bf\Phi_2$, eq.~(\ref{P2eom}), to the boundary, our construction 
is consistent.

%%%%%%%%%%%%%%%%%%%%%%%%%%%%%%%%%%%%%%%%%%%%%%%%%%%%%%%%%%%%%%%
\subsection{OP with singular $\bf\Phi_2$}
%%%%%%%%%%%%%%%%%%%%%%%%%%%%%%%%%%%%%%%%%%%%%%%%%%%%%%%%%%%%%%%

In the orbifold picture, the bulk-plus-brane Lagrangian, corresponding
to the bulk-plus-boundary action (\ref{SJ}), turns out to be
given by
\bea
\label{LG}
\boxed{ \rule[-5pt]{0pt}{18pt} \quad
\mc{L}=\frac{1}{8} {\bf V_2} D\Dbar^2 D {\bf V_2} 
+ \Big[{\bf Z_2} -2{\bf G}\da(z) \Big]^2
+\mc{L}_4\da(z).
\quad}
\eea
The first part of it, explicitly showing $\bf V_2$, is gauge invariant
(up to a total $\p_m$ derivative). As $\bf Z_2$ is gauge invariant,
the brane-localized term $\bf G$ must also be invariant under the
bulk gauge transformation for the Lagrangian to be supersymmetric.

The full square structure of the interaction is required to
guarantee that equations of motion for the bulk and brane
fields are consistent with each other. We have
\bea
\frac{\da\mc{L}}{\da{\bf V_2}} &\equiv&
\frac{1}{4}D\Dbar^2 D {\bf V_2}
-2\p_5 \Big[ {\bf Z_2}-2{\bf G}\da(z) \Big] =0 
\nn\\
\frac{\da\mc{L}}{\da{\bf V}} &\equiv&
\da(z)\left\{
-4\big[ {\bf Z_2}-2 {\bf G}\da(z) \big] 
\frac{\da{\bf G}}{\da{\bf V}}
+\frac{\da{\mc{L}_4}}{\da{\bf V}}
\right\}
=0 \; ,
\eea
so that both equations require $\bf Z_2$ to have the same
singular part,
\bea 
\label{ZGns}
\boxed{ \rule[-5pt]{0pt}{18pt} \quad
{\bf Z_2} 
= 2{\bf G}\da(z) +\text{n.s.} ,
\quad}
\eea
where ``n.s.'' stands for non-singular terms.
As ${\bf Z_2}
=\p_5 {\bf V_2}-\frac{1}{2}({\bf\Phi_2}+{\bf\Phi_2}^\dagger)$,
the singular term can arise from a jump in the odd superfield $\bf V_2$,
\bea
\p_5 {\bf V_2} 
%=\p_5\left[ \ep(z){\bf V_2}^{(+)} \right]
%=\ep^\prime(z){\bf V_2}^{(+)}+\text{n.s.}
=2\da(z){\bf V_2}^{(+)}+\text{n.s.},
\eea
or from the even superfield $\bf \Phi_2$ having a singular part.
If we write
\bea
\label{P2sing}
\boxed{ \rule[-5pt]{0pt}{18pt} \quad
{\bf\Phi_2}={\bf\wt\Phi_2}+4{\bf K}\da(z) ,
\quad}
\eea
with $\bf\wt\Phi_2$ being non-singular, we find that eq.~(\ref{ZGns})
gives rise to a boundary condition
\bea
\label{V2bc}
{\bf V_2}\beq {\bf J} ={\bf K+K^\dagger+G} \; ,
\eea
which coincides exactly with the boundary condition found in the 
boundary picture. Moreover, the gauge transformation of $\bf\Phi_2$,
eq.~(\ref{V2gauge}), when split into the singular and non-singular
parts, gives
\bea
\da_u {\bf K}={\bf\La_2^{(+)}}(u), \quad
\da_u {\bf \wt\Phi_2}=2\p_5{\bf \La_2}(u)-4{\bf\La_2^{(+)}}(u)\da(z) ,
\eea
which implies that the gauge and supersymmetry transformations
of $\bf K$ are exactly as in eq.~(\ref{susyK}). We conclude, therefore,
that the boundary compensator $\bf K$
corresponds to the singular part of $\bf \Phi_2$ in the orbifold
picture.

%%%%%%%%%%%%%%%%%%%%%%%%%%%%%%%%%%%%%%%%%%%%%%%%%%%%%%%%%%%%%%%
\subsection{OP with non-singular $\bf\Phi_2$}
%%%%%%%%%%%%%%%%%%%%%%%%%%%%%%%%%%%%%%%%%%%%%%%%%%%%%%%%%%%%%%%

There is another way to approach bulk-brane coupling in the
orbifold picture. Let us require that $\bf\Phi_2$
be non-singular. This forces us to modify gauge and supersymmetry
transformations of $\bf\Phi_2$ in a way that makes them non-singular,
which gives 
\bea
\label{mP2susy}
\da_u^\prime {\bf \Phi_2} &=& 
2\p_5{\bf \La_2}(u)-4{\bf\La_2^{(+)}}(u)\da(z) \nn\\
\da_\eta^\prime {\bf \Phi_2} &=&
(\eta Q+\etabar\Qbar){\bf \Phi_2}
+2\p_5{\bf \La_2}(\eta)-4{\bf\La_2^{(+)}}(\eta)\da(z) .
\eea
With this modification, $\bf Z_2$ is no longer gauge invariant
\bea
\da_u^\prime {\bf Z_2}=
2\left[ {\bf\La_2}^{(+)}(u)
+\big( {\bf\La_2}^{(+)}(u) \big)^\dagger \right] \da(z) .
\eea
Therefore, the right bulk-plus-brane Lagrangian now is 
\bea
\label{LJ}
\boxed{ \rule[-5pt]{0pt}{18pt} \quad
\mc{L}=\frac{1}{8} {\bf V_2} D\Dbar^2 D {\bf V_2} 
+ \Big[{\bf Z_2} -2{\bf J}\da(z) \Big]^2
+\mc{L}_4\da(z),
\quad}
\eea
where $\bf J$ is required to transform as in eq.~(\ref{susyJ})
in order for $\mc{L}$ to be supersymmetric. As in the boundary
picture, we are lead to $\bf J$ of the form (\ref{JKG}),
explicitly containing the compensator $\bf K$.
Note that, unlike the boundary picture case, we can make a replacement
\bea
\label{LaJ}
{\bf\La_2}^{(+)}(\eta)
\quad \longrightarrow \quad
{\bf\La_J}(\eta)\equiv {\bf\La_2}^{(+)}(\eta)_{\big|
{\bf V_2=J}}
\eea
in the supersymmetry transformations of $\bf J$, $\bf K$,
and $\bf\Phi_2$, and the Lagrangian (\ref{LJ}) would still
be supersymmetric \emph{without} using boundary conditions.

The two orbifold picture constructions are, obviously, related
by the field redefinition (\ref{P2sing}). The advantage of the
formulation with a singular $\bf\Phi_2$ is that it avoids
explicit appearance of the compensator $\bf K$. We will see
more explicitly how the two approaches are related when we
consider the component formulation.

%%%%%%%%%%%%%%%%%%%%%%%%%%%%%%%%%%%%%%%%%%%%%%%%%%%%%%%%%%%%%%%
%%%%%%%%%%%%%%%%%%%%%%%%%%%%%%%%%%%%%%%%%%%%%%%%%%%%%%%%%%%%%%%
\section{Bulk-brane coupling in components}
\label{sec-ccoup}
%%%%%%%%%%%%%%%%%%%%%%%%%%%%%%%%%%%%%%%%%%%%%%%%%%%%%%%%%%%%%%%
%%%%%%%%%%%%%%%%%%%%%%%%%%%%%%%%%%%%%%%%%%%%%%%%%%%%%%%%%%%%%%%

In this section, we show how to go from the superfield 
bulk-brane coupling established in the previous section, 
to its component form. In the boundary picture, we find that
the $Y$-term of ref.~\cite{my2} arises naturally from the extra
superfield boundary term in $S_5^{\prime\prime}$.
In the orbifold picture, we find that in order to arrive at
the form of the coupling established for the Horava-Witten
and Randall-Sundrum scenarios, one has to do a partial
field redefinition.

%%%%%%%%%%%%%%%%%%%%%%%%%%%%%%%%%%%%%%%%%%%%%%%%%%%%%%%%%%%%%%%
\subsection{Boundary conditions}
%%%%%%%%%%%%%%%%%%%%%%%%%%%%%%%%%%%%%%%%%%%%%%%%%%%%%%%%%%%%%%%

In both the boundary and orbifold picture, the boundary condition
is given by eq.~(\ref{V2bc}). As $\bf J$ is a real vector superfield,
we write its components as follows (see Appendix \ref{app1})
\bea
{\bf J} &=& (C_J, \; \chi_J, \; M_J; \; J_m, \; \la_J, \; D_J) .
\eea
With $\bf V_2$ being in the WZ gauge and given by eq.~(\ref{V2WZ}),
the boundary condition (\ref{V2bc}) splits into two sets of
component boundary conditions. The first set requires the three
lowest components of $\bf J$ to vanish:
\bea
\label{cbc1}
C_J=\chi_J=M_J=0 .
\eea
The second set gives the actual boundary conditions in the
component formulation,
\bea
\label{cbc2}
\boxed{ \rule[-5pt]{0pt}{18pt} \quad
A_m \beq J_m, \quad
\la_1 \beq \la_J, \quad
X_3-\p_5\Phi \beq D_J . 
\quad}
\eea

%%%%%%%%%%%%%%%%%%%%%%%%%%%%%%%%%%%%%%%%%%%%%%%%%%%%%%%%%%%%%%%
\subsection{Compensator (super)field}
\label{sec-comp}
%%%%%%%%%%%%%%%%%%%%%%%%%%%%%%%%%%%%%%%%%%%%%%%%%%%%%%%%%%%%%%%

The set of restrictions on $\bf J$, given in eq.~(\ref{cbc1}),
fixes $\bf K$ up to a single real field $K$. To see how this
happens, we first define the components of $\bf G$ and $\bf K$ in
a general way
\bea
\label{Gc}
{\bf G} = (C_G, \; \chi_G, \; M_G; \; G_m, \; \la_G, \; D_G), \quad
{\bf K}=(\phi_K, \; \psi_K, \; F_K).
\eea
Writing $\bf J=K+K^\dagger+G$ in components, we find
\bea
& C_J=\phi_K+\phi_K^\ast+C_G, \quad
\chi_J=-i\st\psi_K+\chi_G, \quad
M_J=-2i F_K+M_G &
\nn\\[5pt]
& J_m= -i\p_m(\phi_K-\phi_K^\ast)+G_m, \quad
\la_J = \la_G, \quad
D_J= D_G . & 
\eea
The restriction (\ref{cbc1}) now gives three equations
on the components of $\bf K$, which leave undetermined only the
imaginary part of its lowest component. Denoting the latter
by~$K$, we have
\bea
\label{defK}
{\bf K} \; = \; \left(-\frac{1}{2}C_G+\frac{i}{2} K, \quad
-\frac{i}{\st}\chi_G, \quad
-\frac{i}{2}M_G \; \right).
\eea
With this definition of $K$, the non-zero components of 
$\bf J$ become
\bea
\label{JGK}
\boxed{ \rule[-5pt]{0pt}{18pt} \quad
J_m=G_m+\p_m K, \quad \la_J=\la_G, \quad D_J=D_G .
\quad}
\eea

Gauge and supersymmetry transformations of the components
of $\bf K$ and $\bf G$ can be found from the superfield
transformations given in eqs.~(\ref{susyK}) and (\ref{susyG}), 
respectively.
(For supersymmetry transformations, eq.~(\ref{bare}) is useful.)
We find, for example,
\bea
& \da_u\phi_K=\dst\frac{i}{2}u^{(+)}, \quad
\da_\eta\phi_K=\st\eta\psi_K & \nn\\[5pt]
& \da_u C_G=0, \quad
\da_\eta C_G=i\eta\chi_G+h.c. &
\eea
Applying these transformations to the lowest component of
eq.~(\ref{defK}), we obtain the following
gauge and supersymmetry transformations of $K$,
\bea
\label{susyKc}
\boxed{ \rule[-5pt]{0pt}{18pt} \quad
\da_u K=u^{(+)}, \quad \da_\eta K=-\eta\chi_G+h.c.
\quad}
\eea
Analogous treatment of the other two components in eq.~(\ref{defK})
reproduces the boundary conditions (\ref{cbc2}) for $A_m$ and $\la_1$.
Note that these boundary conditions would not arise here
if we make the replacement (\ref{LaJ}) in the supersymmetry
transformation of~$\bf K$.

%%%%%%%%%%%%%%%%%%%%%%%%%%%%%%%%%%%%%%%%%%%%%%%%%%%%%%%%%%%%%%%
\subsection{Boundary picture}
%%%%%%%%%%%%%%%%%%%%%%%%%%%%%%%%%%%%%%%%%%%%%%%%%%%%%%%%%%%%%%%

The boundary picture action $S_5^{\prime\prime}$, eq.~(\ref{S5pp}),
appropriate for the odd $A_m$, differs from the original bulk
action $S_5$ by a boundary term that we call $Y$-term \cite{my2,my1},
\bea
S_5^{\prime\prime}=S_5+\int_{\p\mc{M}}Y^{\prime\prime} .
\eea
This $Y^{\prime\prime}$-term is a sum of the $Y^\prime$-term for the 
action $S_5^\prime$, eq.~(\ref{S5p}), and of the boundary superfield term
in eq.~(\ref{S5pp}),
\bea
Y^{\prime\prime} = Y^\prime+2({\bf Z_2 V_2})_{\big|\ta^2\tabar^2}
=F_{m5}A^m-\frac{1}{2}(\la_1\la_2+h.c.) .
\eea
This way we reproduce the $Y$-term of the form 
suggested in ref.~\cite{my2}, with the $F_{m5}A^m$ term present.
For the total bulk-plus-boundary action (\ref{SJ}), we find
\bea
\label{SJY}
S &=& S_5+\dst\int_{\p\mc{M}} \Big[
F_{m5}A^m-\frac{1}{2}(\la_1\la_2+h.c.) \Big] \nn\\[13pt]
&&+\dst\frac{1}{2}\int_{\p\mc{M}} \Big[\mc{L}_4
+2\Phi D_J+2(\la_2\la_J+h.c.)-2F_{m5}J^m \Big] .
\eea
As we will show in Section \ref{sec-bponsh}, this action
is supersymmetric under the bulk supersymmetry transformations
(\ref{bulksusy}) and appropriate transformations of
the components of $\bf J$. We will find, however, 
that showing this requires 
using the boundary condition~(\ref{cbc2}) for~$A_m$
(and also the one for $\la_1$, unless we eliminate auxiliary fields).

We can simplify the form of the action by explicitly using some
or all of the boundary conditions (\ref{cbc2}). Using the one
for $A_m$, we obtain
\bea
\label{S1}
S_1 &=& S_5+\dst\int_{\p\mc{M}} \Big[
-\frac{1}{2}(\la_1\la_2+h.c.) \Big] \nn\\[13pt]
&&+\dst\frac{1}{2}\int_{\p\mc{M}} \Big[\mc{L}_4
+2\Phi D_J+2(\la_2\la_J+h.c.) \Big] .
\eea
Using the boundary conditions for both $A_m$ and $\la_1$, we get
\bea
\label{S2}
S_2=S_5+ \frac{1}{2}\int_{\p\mc{M}} \Big[ \mc{L}_4+
2\Phi D_J+(\la_2\la_J+h.c.) \Big] .
\eea
We will find that supersymmetry of $S_1$
depends on using the boundary conditions for $A_m$ and $\la_1$,
whereas $S_2$ is supersymmetric provided the third boundary
condition in eq.~(\ref{cbc2}) is also used.
The reason for this is explained in Appendix \ref{app3}.

%%%%%%%%%%%%%%%%%%%%%%%%%%%%%%%%%%%%%%%%%%%%%%%%%%%%%%%%%%%%%%%
\subsection{OP with singular fields}
%%%%%%%%%%%%%%%%%%%%%%%%%%%%%%%%%%%%%%%%%%%%%%%%%%%%%%%%%%%%%%%

In the orbifold picture, all $\da(z)$-dependent
terms in the bulk-plus-brane Lagrangian (\ref{LG})
come from the following part 
\bea
\label{Z2Gc}
&&\hspace{-30pt}
\big[ {\bf Z_2}-2{\bf G}\da(z) \big]^2{}_{\big|\ta^2\tabar{}^2} =
\nn\\
&&
-\big[\la_2-2\chi_G\da(z)\big]
 \Big[\p_5\la_1-2\la_G\da(z)
+\frac{i}{2}\si^m\p_m \big[\labar_2-2\chibar_G\da(z)\big]
 \Big]
+h.c. 
\nn\\
&& 
-\big[\Phi+2C_G\da(z)\big]
 \Big[\p_5(X_3-\p_5\Phi)-2D_G\da(z)
-\frac{1}{2}\p_m\p^m\big[\Phi+2C_G\da(z)\big]
 \Big]
\nn\\
&& 
-\frac{1}{2}\big[F_{m5}+2G_m\da(z)\big]^2
+\frac{1}{2}\big[X_{12}-2i M_G\da(z)\big]
            \big[X_{12}^\ast+2i M_G^\ast\da(z)\big].
\eea
Dropping some total $\p_5$ derivatives, irrelevant in the
orbifold picture, the total Lagrangian can be brought 
to the following form
\bea
%\label{LJJ}
\mc{L}=\mc{L}_5+\big[\mc{L}_4+B_1 \big]\da(z)
+B_2\da(z)^2+B_3\da^\prime(z) ,
\eea
where
\bea
B_1 &=& 2\la_2\la_G
+2i\chi_G\si^m\p_m\labar_2
-i X_{12}^\ast M_G
+h.c. \nn\\[5pt]
&&-2F_{m5}G^m+2\Phi D_G
+2 C_G \p_m\p^m\Phi
\nn\\[5pt]
B_2 &=& -4\chi_G\la_G-2i\chi_G\si^m\p_m\chibar_G +h.c. 
\nn\\[5pt]
&& -2 G_m G^m +2C_G\p_m\p^m C_G +4C_G D_G +2 M_G M_G^\ast
\nn\\[5pt]
B_3 &=& -2\chi_G\la_1+h.c.+2C_G(X_3-\p_5\Phi).
\eea
This Lagrangian, by construction, is supersymmetric under the
original supersymmetry transformations (\ref{bulksusy})
of the bulk fields. However, its $\da(z)$-dependent terms 
happen to be more complicated than those
in the (more complicated) supergravity theories. We will see next
that this apparent paradox can be resolved by a simple field
redefinition.

%%%%%%%%%%%%%%%%%%%%%%%%%%%%%%%%%%%%%%%%%%%%%%%%%%%%%%%%%%%%%%%
\subsection{OP with singular $A_5$}
\label{sec-sA5}
%%%%%%%%%%%%%%%%%%%%%%%%%%%%%%%%%%%%%%%%%%%%%%%%%%%%%%%%%%%%%%%

From eq.~(\ref{ZGns}), we know that ${\bf Z_2}-2{\bf G}\da(z)$
is non-singular. Using the component forms of $\bf Z_2$ and
$\bf G$, eqs.~(\ref{Z2c}) and (\ref{Gc}), respectively,
we find that the following fields,
\bea
\label{redef}
\wt\Phi &\equiv& \Phi+2C_G\da(z) \nn\\
\wt\la_2 &\equiv& \la_2-2\chi_G\da(z) \nn\\
\wt X_{12} &\equiv& X_{12}-2i M_G\da(z) \nn\\
\wt X_3 &\equiv& X_3+2C_G\da^\prime(z),
\eea
are non-singular.\footnote{
When we say that a field is non-singular,
we mean that it is non-singular
when equations of motion are used.
Note that we reserve the word ``on-shell'' to mean 
``when auxiliary fields are eliminated.'' 
}
A glance at eq.~(\ref{Z2Gc}) shows that
transforming to the new fields absorbs most of the
$\da(z)$ terms. Performing the field redefinition, and omitting
the tildes, we find
\bea
\label{LGc}
\boxed{ \rule[-10pt]{0pt}{25pt} \quad
\mc{L}=\mc{L}_5^{(\mc{F})}
+\Big[ \mc{L}_4+2\Phi D_G+2(\la_2\la_G+h.c.) \Big]\da(z) ,
\quad}
\eea
where $\mc{L}_5^{(\mc{F})}$ is obtained from the original
Lagrangian $\mc{L}_5$, eq.~(\ref{L5mn}), by replacing $F_{m5}$ with
\bea
\label{FG}
\mc{F}_{m5}= F_{m5}+2G_m\da(z) .
\eea
Performing the redefinition (\ref{redef}) 
in the supersymmetry transformations (\ref{bulksusy})
requires using the
transformations of the components of $\bf G$. Since $\bf G$
transforms as in eq.~(\ref{susyG}), its components transform
according to eq.~(\ref{bare}). After a short calculation, we find
the following modified supersymmetry transformations of the bulk fields,
\bea
%\label{bulksusy}
\da_\eta A_m &=& i \eta\si_m\labar_1+h.c. \nn\\[5pt]
\da_\eta A_5 &=& -\eta\la_2 -2\eta\chi_G\da(z) +h.c. \nn\\[5pt]
\da_\eta\Phi &=& -i\eta\la_2+h.c. \nn\\[5pt]
%%%%%
\da_\eta\la_1 &=& \si^{mn}\eta F_{mn}+i(X_3-\p_5\Phi)\eta 
\nn\\[5pt]
\da_\eta\la_2 &=& -i\si^m\etabar \big[ F_{m5}+2G_m\da(z) \big]
-\si^m\etabar\p_m\Phi-i X_{12}\eta
\nn\\[5pt]
%%%%%
\da_\eta X_{12} &=& 
2i\etabar \big[\p_5\labar_1-2\la_G\da(z) \big]
-2\etabar\sibar^m\p_m\la_2
\nn\\[5pt]
\da_\eta X_3 &=& 
-i\eta\p_5\la_2-\eta\si^m\p_m\labar_1
+h.c. 
\eea
The modifications can be summarized as follows:
1) replace $F_{m5}$ with $\mc{F}_{m5}$,
2) modify the transformation of $A_5$ 
by adding the following singular piece
\bea
\label{A5ss}
\da_\eta^{\rm (s)} A_5 &=& -2(\eta\chi_G+h.c.)\da(z) , 
\eea
and 3) modify the transformation of $X_{12}$ (the even auxiliary field)
by terms that make it non-singular when the boundary conditions
(\ref{cbc2}) are used. When auxiliary fields are eliminated, 
we need only the first two prescriptions. 
Therefore, in the on-shell formulation, we match the
supergravity bulk-brane coupling construction of ref.~\cite{falk}.

Note that after the redefinition (\ref{redef}), we still have
one singular field left: $A_5$. From eq.~(\ref{ZGns}) and the
boundary conditions (\ref{cbc2}), we have
\bea
F_{m5}+2G_m\da(z)=n.s., \quad
A_m \beq G_m+\p_m K 
\quad \Rightarrow \quad
\boxed{ \rule[-5pt]{0pt}{18pt} \quad
A_5=2K\da(z)+n.s.
\quad}
\eea
We see that the singular part of $A_5$ is directly related to the
compensator field $K$. If we redefine $A_5$ to make it non-singular, 
we find that its supersymmetry transformation also becomes non-singular:
\bea
\label{A5redef}
\wt A_5=A_5-2K\da(z) \quad \Rightarrow \quad
\da_\eta \wt A_5 = -\eta\la_2+h.c.
\eea
If we now replace $A_5$ with $\wt A_5$ in the expression for
$\mc{F}_{m5}$, eq.~(\ref{FG}), we find that $G_m$ gets replaced
by $J_m=G_m+\p_m K$:
\bea
\boxed{ \rule[-5pt]{0pt}{18pt} \quad
\mc{F}_{m5} 
= F_{m5}+2G_m\da(z)
= \wt{F}_{m5}+2J_m\da(z).
\quad}
\eea
As we will see next, after this final field redefinition
we come exactly to the construction in which the superfield
$\bf\Phi_2$ is non-singular from the start.

%%%%%%%%%%%%%%%%%%%%%%%%%%%%%%%%%%%%%%%%%%%%%%%%%%%%%%%%%%%%%%%
\subsection{OP with non-singular fields}
%%%%%%%%%%%%%%%%%%%%%%%%%%%%%%%%%%%%%%%%%%%%%%%%%%%%%%%%%%%%%%%

In the case with non-singular $\bf \Phi_2$, the bulk-plus-brane
Lagrangian is given by eq.~(\ref{LJ}). As the lowest components
of $\bf J$ (unlike $\bf G$) vanish, $C_J=\chi_J=M_J=0$, the
component form of the Lagrangian is simple without any field
redefinitions:
\bea
\label{LJc}
\mc{L}=\mc{L}_5^{(\mc{F})}
+\Big[ \mc{L}_4+2\Phi D_J+2(\la_2\la_J+h.c.) \Big]\da(z).
\eea
As before, we must replace $F_{m5}$ by $\mc{F}_{m5}$ that
is now given by
\bea
\mc{F}_{m5}= F_{m5}+2J_m\da(z).
\eea
Superfield supersymmetry transformations are now different from
those in eq.~(\ref{V2susy}). They are modified as in 
eq.~(\ref{mP2susy}) so that the transformation of $\bf \Phi_2$
is non-singular. We should, however, make the choice: whether
or not to make the replacement (\ref{LaJ}). Because of the last
statement in Section \ref{sec-comp}, the component Lagrangian
will be supersymmetric \emph{without} using boundary conditions provided
we do make the replacement (\ref{LaJ}). The component supersymmetry
transformations then become
\bea
%\label{bulksusy}
\da_\eta A_m &=& i \eta\si_m\labar_1+h.c. \nn\\[5pt]
\da_\eta A_5 &=& -\eta\la_2 +h.c. \nn\\[5pt]
\da_\eta\Phi &=& -i\eta\la_2+h.c. \nn\\[5pt]
%%%%%
\da_\eta\la_1 &=& \si^{mn}\eta F_{mn}+i(X_3-\p_5\Phi)\eta 
\nn\\[5pt]
\da_\eta\la_2 &=& -i\si^m\etabar \big[ F_{m5}+2J_m\da(z) \big]
-\si^m\etabar\p_m\Phi-i X_{12}\eta
\nn\\[5pt]
%%%%%
\da_\eta X_{12} &=& 
2i\etabar \big[\p_5\labar_1-2\la_J\da(z) \big]
-2\etabar\sibar^m\p_m\la_2
\nn\\[5pt]
\da_\eta X_3 &=& 
-i\eta\p_5\la_2-\eta\si^m\p_m\labar_1
+h.c. 
\eea
This differs from the original transformations (\ref{bulksusy})
by $\da(z)$-dependent modifications that
%The $\da(z)$-dependent modifications, compared to eq.~(\ref{bulksusy}),
are now all covered by one simple rule \cite{bb1, my2, my1}: 
the modifications must
make the transformations non-singular when the boundary conditions 
are used.

We conclude that there are two alternative simple forms of the
bulk-brane coupling in the orbifold picture: one with the compensator
$K$ appearing explicitly via $J_m$, 
and the other where the role of the compensator
is played by the singular part of $A_5$. 
The two formulations are related
to each other by the redefinition (\ref{A5redef}) of $A_5$.

%%%%%%%%%%%%%%%%%%%%%%%%%%%%%%%%%%%%%%%%%%%%%%%%%%%%%%%%%%%%%%%
%%%%%%%%%%%%%%%%%%%%%%%%%%%%%%%%%%%%%%%%%%%%%%%%%%%%%%%%%%%%%%%
\section{On-shell coupling}
\label{sec-onsh}
%%%%%%%%%%%%%%%%%%%%%%%%%%%%%%%%%%%%%%%%%%%%%%%%%%%%%%%%%%%%%%%
%%%%%%%%%%%%%%%%%%%%%%%%%%%%%%%%%%%%%%%%%%%%%%%%%%%%%%%%%%%%%%%

In this section, we go on-shell (eliminate auxiliary fields)
and check explicitly that the bulk-plus-brane/boundary actions
we constructed are indeed supersymmetric. We find that some
boundary conditions have to be used for supersymmetry in the
boundary picture. At the end of the section,
we give an explicit example of a coupled bulk-brane system
which makes contact with the supergravity construction of 
ref.~\cite{falk}.

%%%%%%%%%%%%%%%%%%%%%%%%%%%%%%%%%%%%%%%%%%%%%%%%%%%%%%%%%%%%%%%
\subsection{Modified Bianchi identity}
%%%%%%%%%%%%%%%%%%%%%%%%%%%%%%%%%%%%%%%%%%%%%%%%%%%%%%%%%%%%%%%

As we established, in the orbifold picture, a part of the bulk-brane
coupling prescription consists in replacing $F_{m5}$ with $\mc{F}_{m5}$
both in the Lagrangian and in the supersymmetry transformations.
Let us now generalize this to the following shift
\bea
\label{BMN}
F_{MN}\equiv \p_M A_N-\p_N A_M
\quad \longrightarrow \quad
\mc{F}_{MN}=F_{MN}+B_{MN} .
\eea
On-shell ($X_a=0$) and after the shift,
the bulk Lagrangian (\ref{L5}) turns into
\bea
%\label{L5}
\mc{L}_5^{(\mc{F})}=
-\frac{1}{4} \mc{F}_{MN} \mc{F}^{MN}-\frac{1}{2}\p_M\Phi\p^M\Phi
-\frac{i}{2}\wt\La^i\Ga^M\p_M\La_i \; ,
\eea
and the corresponding supersymmetry transformations become
\bea
%\label{MPsusytr}
\da_\Eta A_M &=& i\wt\Eta^i\Ga_M\La_i \nn\\
\da_\Eta\Phi &=& i\wt\Eta^i\La_i \nn\\
\da_\Eta\La_i &=& (\Si^{MN}\mc{F}_{MN}+\Ga^M\p_M\Phi)\Eta_i .
\eea
Supersymmetry transformation of the bulk Lagrangian now
produces not only the total derivative, but also extra terms
involving $B_{MN}$:
\bea
\label{modB}
\da_\Eta\mc{L}_5 &=&
\p_M \wt K^M  
-\frac{1}{2}\mc{F}^{MN}\da_\Eta B_{MN}
-\frac{i}{2}\wt\eta^i\Ga^{MNK}\La_i\p_K\mc{F}_{MN} \nn\\
&&
\wt K^M=
-\mc{F}^{MN}\da_\Eta A_N-\da_\Eta\Phi\p^M\Phi
+\frac{i}{2}\wt\La^i\Ga^M\da_\Eta\La_i.
\eea
The last term in $\da_\Eta\mc{L}_5$ is the famous
contribution due to the ``modified Bianchi identity.''
Note that in the boundary picture, we have $B_{MN}=0$
and the total derivative term is important; 
in the orbifold picture, $B_{MN}\neq 0$ and
the total derivative is irrelevant.

%%%%%%%%%%%%%%%%%%%%%%%%%%%%%%%%%%%%%%%%%%%%%%%%%%%%%%%%%%%%%%%
\subsection{Boundary picture}
\label{sec-bponsh}
%%%%%%%%%%%%%%%%%%%%%%%%%%%%%%%%%%%%%%%%%%%%%%%%%%%%%%%%%%%%%%%

The bulk-plus-boundary action in the boundary picture is given
by eq.~(\ref{SJ}),
\bea
%\label{SJY}
S &=& S_5+\dst\int_{\p\mc{M}} \Big[
F_{m5}A^m-\frac{1}{2}(\la_1\la_2+h.c.) \Big] \nn\\[13pt]
&&+\dst\frac{1}{2}\int_{\p\mc{M}} \Big[\mc{L}_4
+2\Phi D_J+2(\la_2\la_J+h.c.)-2F_{m5}J^m \Big] .
\eea
Supersymmetry variation of $S_5$ produces the following 
boundary term, 
\bea
\da_\eta S_5
=\int_{\p\mc{M}}(-\wt K^5)
=\int_{\p\mc{M}}\big[ 
-F^{m5}\da_\eta A_m
+\da_\eta\Phi\p_5\Phi
-\frac{1}{2}(\la_2\da_\eta\la_1-\la_1\da_\eta\la_2+h.c.)
\big] .
\eea
To find the variation of the total action, we need
to know supersymmetry transformations of $J_m$, $\la_J$, and $D_J$.
We know that components of $\bf G$ transform as in eq.~(\ref{bare}),
so that, in particular,~\footnote{
In our notation, 
$v_{mn}=\p_m v_n-\p_n v_m$,
$G_{mn}=\p_m G_n-\p_n G_m$,
$J_{mn}=\p_m J_n-\p_n J_m$.
}
\bea
\da_\eta G_m &=& i\eta\si_m\labar_G+\p_m(\eta\chi_G)+h.c. \nn\\
\da_\eta \la_G &=& \si^{mn}\eta G_{mn}+i\eta D_G \nn\\
\da_\eta D_G &=& \etabar\sibar^m\p_m\la_G+h.c.
\eea
From eq.~(\ref{JGK}) and the supersymmetry transformation
(\ref{susyKc}) of $K$, it then follows that
\bea
\da_\eta J_m &=& i\eta\si_m\labar_J+h.c. \nn\\
\da_\eta \la_J &=& \si^{mn}\eta J_{mn}+i\eta D_J \nn\\
\da_\eta D_J &=& \etabar\sibar^m\p_m\la_J+h.c.
\eea
Using these transformations together with the ones for the
bulk fields, eq.~(\ref{bulksusy}), we find
\bea
\label{susyS}
\da_\eta S &=& \int_{\p\mc{M}} \Big[
\eta\si^{mn}\la_2(F_{mn}-J_{mn})+h.c.
+(A^m-J^m)\da_\eta F_{m5} \Big].
\eea
For the action (\ref{S1}), obtained from $S$ by 
using the $A_m$ boundary condition, we have
\bea
\da_\eta S_1 &=& \int_{\p\mc{M}} \Big[
\eta\si^{mn}\la_2(F_{mn}-J_{mn})
-i\eta\si^m(\labar_1-\labar_J)F_{m5}
+h.c. \Big] .
\eea
For the action (\ref{S2}), obtained from
$S_1$ by using the $\la_1$ boundary condition, we get
\bea
\da_\eta S_2 &=& \int_{\p\mc{M}} \Big[
\frac{1}{2}\eta\si^{mn}\la_2(F_{mn}-J_{mn})
-\frac{i}{2}\eta\si^m(\labar_1-\labar_J)F_{m5} \nn\\
&&\hspace{30pt} -\frac{i}{2}\eta\la_2(\p_5\Phi+D_J)
-\frac{1}{2}\eta\si^m(\labar_1-\labar_J)\p_m\Phi
+h.c. \Big] .
\eea
We conclude that each action is supersymmetric, and in each
case supersymmetry of the action
depends on using some boundary conditions.
The basic pattern we observe is: the more boundary conditions are
used to simplify the action, the more of them
are needed to prove its supersymmetry.
The way to predict which boundary conditions are needed
in each case is given in Appendix \ref{app3}.

%%%%%%%%%%%%%%%%%%%%%%%%%%%%%%%%%%%%%%%%%%%%%%%%%%%%%%%%%%%%%%%
\subsection{Orbifold picture}
%%%%%%%%%%%%%%%%%%%%%%%%%%%%%%%%%%%%%%%%%%%%%%%%%%%%%%%%%%%%%%%

In the orbifold picture, with singular $A_5$, we have 
\bea
& B_{mn}=0, \quad B_{m5}=-B_{5m}=2G_m\da(z) & \nn\\[5pt]
& \mc{F}_{mn}=F_{mn}, \quad
\mc{F}_{m5}=F_{m5}+2G_m\da(z). &
\eea
The bulk-plus-brane Lagrangian is given by eq.~(\ref{LGc}),
\bea
\label{L4pr}
\mc{L}=\mc{L}_5^{(\mc{F})}+\mc{L}_4^\prime\da(z), \quad
\mc{L}_4^\prime= \mc{L}_4+2\Phi D_G+2(\la_2\la_G+h.c.).
\eea
Supersymmetry variation of $\mc{L}_5^{(\mc{F})}$ gives
\bea
\da_\eta \mc{L}_5^{(\mc{F})} = 
\Big\{2(\eta\si^{mn}\la_2+h.c.)G_{mn}
-\mc{F}^{m5} \Big[2 \da_\eta G_m +\p_m(\wt\da_\eta^{(s)} A_5)\Big]
\Big\}\da(z),
\eea
where 
the terms with $G_m$ follow from the $B_{MN}$ terms 
in eq.~(\ref{modB}), and
the last term follows from the modification (\ref{A5ss})
in the supersymmetry transformation of $A_5$ with
\bea
\wt\da_\eta^{(s)} A_5 \equiv -2\eta\chi_G+h.c. =2\da_\eta K .
\eea
Note that the sum of the terms in the square bracket gives
$2\da_\eta J_m$. For $\mc{L}_4^\prime$, we find
\bea
\da_\eta(\la_2\la_G+h.c.+\Phi D_G)=
-\eta\si^{mn}\la_2 G_{mn}
+i\eta\si^m\labar_G \mc{F}_{m5} +h.c.,
\eea
from which we conclude that the total Lagrangian $\mc{L}$ 
is supersymmetric, $\da_\eta \mc{L}=0$,
without using any boundary conditions.

%%%%%%%%%%%%%%%%%%%%%%%%%%%%%%%%%%%%%%%%%%%%%%%%%%%%%%%%%%%%%%%
\subsection{Example}
\label{sec-ex}
%%%%%%%%%%%%%%%%%%%%%%%%%%%%%%%%%%%%%%%%%%%%%%%%%%%%%%%%%%%%%%%

To make contact with the supergravity construction of
ref.~\cite{falk},
we consider an example with one brane-localized chiral 
superfield $\bf \Phi$ and 
\bea
{\bf G =\Phi^\dagger \Phi}.
\eea
With ${\bf\Phi}=(\phi, \psi, F)$, the components of $\bf G$
are given by
\bea
C_G &=& \phi\phi^\ast \nn\\
\chi_G &=& -i\st\phi^\ast\psi \nn\\
M_G &=& -2i F\phi^\ast \nn\\
G_m &=& i(\phi\p_m\phi^\ast-\phi^\ast\p_m\phi)+\psi\si_m\psibar \nn\\
\la_G &=& \st\si^m\psibar\p_m\phi +i\st\psi F^\ast \nn\\
D_G &=& 2F F^\ast-2\p_m\phi\p^m\phi^\ast
- \big( i\psi\si^m\p_m\psibar+h.c. \big).
\eea
The supersymmetry transformation of the compensator $K$,
eq.~(\ref{susyKc}), can now be written as follows,~\footnote{
Note that this form of $\da_\eta K$ implies 
$\da_\eta^{(s)} A_5=
2i(\phi^\ast\da_\eta\phi-\phi\da_\eta\phi^\ast)\da(z)$,
which is remarkably similar to $\wt\da C_{11 AB}$ in 
eq.~(2.16) of the first paper in ref.~\cite{hw}.
}
\bea
\da_\eta K=i\st\phi^\ast\eta\psi+h.c.
=i(\phi^\ast\da_\eta\phi-\phi\da_\eta\phi^\ast),
\eea
which clearly shows that we cannot ``gauge fix'' the compensator
by making it a function of the matter fields.
To complete the setup, we choose 
\bea
\mc{L}_4=\int d^2\ta d^2\tabar {\bf\Phi^\dagger\Phi}
=F F^\ast-\p_m\phi\p^m\phi^\ast
- \left(\frac{i}{2}\psi\si^m\p_m\psibar+h.c. \right).
\eea
Plugging all the pieces into the bulk-plus-brane Lagrangian
(\ref{L4pr}), and eliminating the auxiliary field $F$ by
its equation of motion,
\bea
F=-2i\st(1+4\Phi)^{-1}\la_2\psi,
\eea
we find that the on-shell Lagrangian is given by
$\mc{L}=\mc{L}_5^{(\mc{F})}+\mc{L}_4^\prime\da(z)$ with
\bea
\label{exL4pr}
\mc{L}_4^\prime &=& (1+4\Phi)\left[
-\p_m\phi\p^m\phi^\ast
- \left(\frac{i}{2}\psi\si^m\p_m\psibar+h.c. \right)
\right] \nn\\[5pt]
&& +2\st(\la_2\si^m\psibar\p_m\phi+h.c.) 
-8(1+4\Phi)^{-1}(\la_2\psi)(\labar_2\psibar).
\eea
The bulk Lagrangian $\mc{L}_5^{(\mc{F})}$ is obtained
from $\mc{L}_5$ by replacing $F_{m5}$ with
\bea
\mc{F}_{m5}=F_{m5}+2G_m\da(z), \quad
G_m = i(\phi\p_m\phi^\ast-\phi^\ast\p_m\phi)+\psi\si_m\psibar.
\eea
The same substitution must be made in the supersymmetry transformations
(\ref{bulksusy}), and, in addition, the transformation of $A_5$
should be modified by adding 
\bea
\da^{(s)}A_5=2(\da_\eta K)\da(z)
=2i\st\phi^\ast\eta\psi\da(z)+h.c.
\eea
With these modifications, the total Lagrangian $\mc{L}$ is supersymmetric
without using any boundary conditions.

We observe that the whole construction is identical to the one 
in supergravity \cite{falk}. 
It is also amusing to note that the brane Lagrangian (\ref{exL4pr})
appears to be very similar to the one in supergravity:
$(1+4\Phi)$ plays the role of the induced metric, $\la_2$ is the
``gravitino,'' $\la_2\si^m\psibar\p_m\phi$ is the ``Noether coupling''
term, and $(\la_2\psi)(\labar_2\psibar)$ represents 4-Fermi terms.

%%%%%%%%%%%%%%%%%%%%%%%%%%%%%%%%%%%%%%%%%%%%%%%%%%%%%%%%%%%%%%%
%\newpage
\section{Summary and Conclusions}
\label{sec-sum}
%%%%%%%%%%%%%%%%%%%%%%%%%%%%%%%%%%%%%%%%%%%%%%%%%%%%%%%%%%%%%%%

In this paper, we showed that the basic features of the
supergravity bulk-brane coupling, present both in the 
Horava-Witten (11D) and Randall-Sundrum (5D) scenarios,
appear also in the simplified globally supersymmetric model
we considered (Mirabelli-Peskin model with odd $A_m$).
Using the 4D $N=1$ superfield formulation of the model, we showed
that the full square structure \cite{hw} of the coupling 
in the orbifold picture is present already on the superfield
level (see eq.~(\ref{LG})). In transition to the component
formulation, one has to make some field redefinitions 
(see eq.~(\ref{redef})) in order to arrive at the established
form of the coupling. After the redefinition, the full square
structure remains only for the $(F_{m5}+2G_m\da(z))^2$ term in the
Lagrangian (\ref{LGc}). As the redefined fields are non-singular,
the shift $F_{m5}\;\rightarrow F_{m5}+2G_m\da(z)$ in supersymmetry
transformations is required to make the transformations non-singular.
All together, we recover the ``modified Bianchi identity''
prescription for the coupling \cite{hw}.

The only modification of the supersymmetry transformations,
in the formulation of refs.~\cite{hw,falk},
that is not covered by the prescription ``make them non-singular''
\cite{bb1,my2,my1}
concerns the ``orthogonal'' component of the bulk gauge field.
In fact, $A_5$ is the only field in this formulation which is
singular. We showed that there is another formulation, where
all fields are non-singular, and where the singular part of $A_5$
is replaced by a compensator field $K$.
All modifications of the supersymmetry transformations are
then covered by one simple rule.

Although optional in the orbifold picture,
the presence of the compensator $K$ is unavoidable in the
boundary picture construction. In both pictures, the boundary
condition for the odd gauge field $A_m$ is $A_m\beq J_m=G_m+\p_m K$.
The gauge transformation of $K$, given in eq.~(\ref{susyKc})
(compare also with eq.~(13.7) of ref.~\cite{my2}), guarantees
gauge invariance of the boundary condition.
On the other hand, its supersymmetry transformation (\ref{susyKc})
is such that $K$ together with $C_G$, $\chi_G$, and $M_G$
(the lowest components of $\bf G$) combine into one chiral superfield 
(the compensator superfield $\bf K$, eq.~(\ref{defK})).

Our results shed some more light on the general structure of the
supersymmetric bulk-brane coupling. They should also be useful
in obtaining a more explicit (component) form of the coupling in the
supersymmetric Randall-Sundrum scenario starting from the superfield
formulation developed in ref.~\cite{sfsg}.

\vspace{20pt}
{\bf Acknowledgments.}
I would like to thank Hyun Min Lee and Christoph Ludeling
for a discussion of an earlier version of this work.

%%%%%%%%%%%%%%%%%%%%%%%%%%%%%%%%%%%%%%%%%%%%%%%%%%%%%%%%%%%%%%%
%%%%%%%%%%%%%%%%%%%%%%%%%%%%%%%%%%%%%%%%%%%%%%%%%%%%%%%%%%%%%%%
%\newpage
\section{Appendix}
 \def\theequation{\thesubsection.\arabic{equation}}
 \setcounter{equation}{0}
 \def\thesubsection{A}
%\appendix
\subsection{Superfield components}
\label{app1}
%%%%%%%%%%%%%%%%%%%%%%%%%%%%%%%%%%%%%%%%%%%%%%%%%%%%%%%%%%%%%%%
%%%%%%%%%%%%%%%%%%%%%%%%%%%%%%%%%%%%%%%%%%%%%%%%%%%%%%%%%%%%%%%

Our supersymmetry conventions follow closely those of ref.~\cite{wb}.
For real vector and chiral superfields, we use the following
shorthand notation,
\bea
{\bf V}=(C, \; \chi, \; M; \; v_m, \; \la, \; D), \quad
{\bf \Phi}=(\phi, \; \psi, \; F),
\eea
corresponding to the standard component expansions,
\bea
{\bf V} &=& i\ta\chi+\frac{i}{2}\ta^2 M
-i\tabar{}^2\ta
\big[\la+\frac{i}{2}\si^m\p_m\chibar\big]+h.c. \nn\\
&&+C-\ta\si^m\tabar v_m
+\frac{1}{2}\ta^2\tabar{}^2\big[D+\frac{1}{2}\p_m\p^m C\big] \nn\\
%%%%%
{\bf \Phi} &=& \Big(1+i\ta\si^m\tabar\p_m
                +\frac{1}{4}\ta^2\tabar{}^2\p_m\p^m\Big)\phi
+\st\Big(\ta+\frac{i}{2}\ta^2\tabar\sibar^m\p_m\Big)\psi
+\ta^2 F \; .
\eea
When supersymmetry transformations have the standard form
(without additional gauge transformations),
\bea
\da_\eta {\bf V} =(\eta Q+\etabar\Qbar) {\bf V}, \quad
\da_\eta {\bf \Phi} =(\eta Q+\etabar\Qbar) {\bf \Phi},
\eea
the component transformations are as follows,
\bea
\label{bare}
\da_\eta C &=& i\eta\chi+h.c. \nn\\
\da_\eta \chi &=& \si^m\etabar(\p_m C+i v_m)+\eta M \nn\\
\da_\eta M &=& 2\etabar(\labar+i\sibar^m\p_m\chi) \nn\\
\da_\eta v_m &=& i\eta\si_m\labar+\p_m(\eta\chi)+h.c \nn\\
\da_\eta \la &=& \si^{mn}\eta v_{mn}+i\eta D \nn\\
\da_\eta D &=& \p_m(\etabar\sibar^m\la+h.c.) \nn\\
\da_\eta \phi &=& \st\eta\psi \nn\\
\da_\eta \psi &=& i\st\si^m\etabar\p_m\phi+\st\eta F \nn\\
\da_\eta F &=& \p_m(i\st\etabar\sibar^m\psi).
\eea
Components of gauge invariant superfields, 
$\bf Z_2$, $\bf W_2$, and $\bf G$,
have exactly this form of supersymmetry transformations.

%%%%%%%%%%%%%%%%%%%%%%%%%%%%%%%%%%%%%%%%%%%%%%%%%%%%%%%%%%%%%%%
%%%%%%%%%%%%%%%%%%%%%%%%%%%%%%%%%%%%%%%%%%%%%%%%%%%%%%%%%%%%%%%
 \def\theequation{\thesubsection.\arabic{equation}}
 \setcounter{equation}{0}
 \def\thesubsection{B}
\subsection{$\bf V_2$ and $\bf\Phi_2$ without WZ}
\label{app2}
%%%%%%%%%%%%%%%%%%%%%%%%%%%%%%%%%%%%%%%%%%%%%%%%%%%%%%%%%%%%%%%
%%%%%%%%%%%%%%%%%%%%%%%%%%%%%%%%%%%%%%%%%%%%%%%%%%%%%%%%%%%%%%%

Our bulk superfields transform under supersymmetry according
to eq.~(\ref{V2susy}),
\bea
\da_\eta {\bf V_2} &=& (\eta Q+\etabar\Qbar) {\bf V_2}
+{\bf\La_2}(\eta)+{\bf\La_2}(\eta)^\dagger \nn\\
\da_\eta {\bf \Phi_2} &=& (\eta Q+\etabar\Qbar) {\bf \Phi_2}
+2\p_5{\bf\La_2}(\eta) .
\eea
Let us proceed without imposing the WZ gauge. If we take the
gauge parameter in its general form,
${\bf\La_2}(\eta)=(a_2, \al_2, f_2)$ ,
the component transformations become
\bea
\da_\eta C_2 &=& i\eta\chi_2
+a_2+h.c. \nn\\
\da_\eta \chi_2 &=& \si^m\etabar(\p_m C_2+i v_m^{(2)})+\eta M_2
-i\st\al_2 \nn\\
\da_\eta M_2 &=& 2\etabar(\labar_{(2)}+i\sibar^m\p_m\chi_2)
-2i f_2 \nn\\
\da_\eta v_m^{(2)} &=& i\eta\si_m\labar_{(2)}
+\p_m(\eta\chi_2-i a_2)+h.c \nn\\
\da_\eta \la_{(2)} &=& \si^{mn}\eta v_{mn}^{(2)}+i\eta D_2 \nn\\
\da_\eta D_2 &=& \p_m(\etabar\sibar^m\la_{(2)}+h.c.) \nn\\
\da_\eta \phi_2 &=& \st\eta\psi_2 +2\p_5 a_2\nn\\
\da_\eta \psi_2 &=& i\st\si^m\etabar\p_m\phi_2+\st\eta F_2 +2\p_5 \al_2\nn\\
\da_\eta F_2 &=& \p_m(i\st\etabar\sibar^m\psi_2) +2\p_5 f_2 .
\eea
One can check that on fields defined in the following way,
\bea
%\label{bfSF}
& A_m=v_m^{(2)}, \quad 
\Phi+i A_5=-\p_5 C_2+\phi_2 & \nn\\[5pt]
& \la_1=\la_{(2)}, \quad 
\la_2=\p_5\chi_2+\frac{i}{\st}\psi_2 & \nn\\[5pt]
& X_3-\p_5\Phi=D_2, \quad
X_{12}=i\p_5 M_2-F_2 , &
\eea
the supersymmetry transformations take the form 
\bea
%\label{bulksusy}
\da_\eta A_m &=& i \eta\si_m\labar_1+h.c.+\p_m u(\eta) \nn\\[5pt]
\da_\eta A_5 &=& -\eta\la_2+h.c. +\p_5 u(\eta) \nn\\[5pt]
\da_\eta\Phi &=& -i\eta\la_2+h.c. \nn\\[5pt]
%%%%%
\da_\eta\la_1 &=& \si^{mn}\eta F_{mn}+i(X_3-\p_5\Phi)\eta 
\nn\\[5pt]
\da_\eta\la_2 &=& -i\si^m\etabar F_{m5}
-\si^m\etabar\p_m\Phi-i X_{12}\eta
\nn\\[5pt]
%%%%%
\da_\eta X_{12} &=& 
2i\etabar \p_5\labar_1
-2\etabar\sibar^m\p_m\la_2
\nn\\[5pt]
\da_\eta X_3 &=& 
-i\eta\p_5\la_2-\eta\si^m\p_m\labar_1
+h.c. ,
\eea
which differs from eq.~(\ref{bulksusy}) only by the $U(1)$ 
gauge transformation (\ref{U1gtr}) with 
\bea
u(\eta)=\eta\chi_2+\etabar\chibar_2
+2\,{\rm Im}(a_2).
\eea
(Note that ${\rm Re}(a_2)$, $\al_2$, and $f_2$
affect only the transformations of $C_2$, $\chi_2$, and $M_2$.)
If it is required that we stay with
the original field content, then the explicit appearance of $\chi_2$
is a problem. To deal with it, we can choose the extra superfield 
gauge transformation in a way that removes $u(\eta)$.
The simplest choice that does this is given by
\bea
%\label{choice1}
{\bf\La_2}(\eta)=
\left( -\frac{i}{2}(\eta\chi_2+\etabar\chibar_2), \quad
0, \quad 0 \right).
\eea
It affects supersymmetry transformations of only 
$A_m$ and $A_5$,
while leaving those of $C_2$, $\chi_2$ and $M_2$ unchanged.
A more involved choice,
\bea
%\label{choice2}
{\bf\La_2}(\eta)= \left( -i\eta\chi_2, \quad 
-\frac{i}{\st}\Big[ \si^m\etabar(\p_m C_2+i A_m)+\eta M \Big], \quad
-i\etabar\labar_1+\etabar\sibar^m\p_m\chi_2 \right),
\eea
makes $C_2$, $\chi_2$ and $M_2$ supersymmetry invariant. Therefore,
it allows setting these fields to zero, which would put us into 
the Wess-Zumino gauge, $C_2=\chi_2=M_2=0$, 
while turning ${\bf\La_2}(\eta)$ into the
compensating gauge transformation (\ref{WZcomp}).

To summarize, the superfield description 
of the bulk 5D vector multiplet uses two 4D $N=1$ superfields
with the following components,~\footnote{
This form of $\bf V_2$ and $\bf\Phi_2$ can be obtained from
eq.~(\ref{V2WZ}) by a gauge transformation with the following
parameter:
${\bf \La_2}=\Big( \frac{1}{2} C_2, \;
                  \frac{i}{\st}\chi_2, \; \frac{i}{2}M_2 \Big)$.
}
\bea
\label{newmap}
&{\bf V_2}=(C_2, \quad \chi_2, \quad M_2; \quad 
           A_m, \quad \la_1, \quad X_3-\p_5\Phi) &
\nn\\[5pt]
& {\bf\Phi_2}=(\Phi+i A_5+\p_5 C_2, \quad 
-i\st(\la_2-\p_5\chi_2), \quad
-X_{12}+i\p_5 M_2) . &
\eea
If the superfield supersymmetry transformations do not involve 
a $\bf\La_2(\eta)$ gauge 
transformation, the component supersymmetry transformations
differ from the ones in eq.~(\ref{bulksusy}) by a $\chi_2$-dependent
$U(1)$ gauge transformation. The latter can be eliminated by
a proper choice of $\bf\La_2(\eta)$. Imposing the WZ gauge 
corresponds to just one of many possible choices.

%%%%%%%%%%%%%%%%%%%%%%%%%%%%%%%%%%%%%%%%%%%%%%%%%%%%%%%%%%%%%%%
%%%%%%%%%%%%%%%%%%%%%%%%%%%%%%%%%%%%%%%%%%%%%%%%%%%%%%%%%%%%%%%
 \def\theequation{\thesubsection.\arabic{equation}}
 \setcounter{equation}{0}
 \def\thesubsection{C}
\subsection{Boundary conditions for supersymmetry}
\label{app3}
%%%%%%%%%%%%%%%%%%%%%%%%%%%%%%%%%%%%%%%%%%%%%%%%%%%%%%%%%%%%%%%
%%%%%%%%%%%%%%%%%%%%%%%%%%%%%%%%%%%%%%%%%%%%%%%%%%%%%%%%%%%%%%%

Deriving the component form of the bulk-plus-boundary action
(\ref{SJ}), with $\bf V_2$ and $\bf\Phi_2$ as in eq.~(\ref{newmap}),
we encounter the following terms,
\bea
%\label{ZVJlong}
2{\bf Z_2(V_2-J)}_{\big|\ta^2\tabar{}^2} &=&
-\, \Phi(X_3-\p_5\Phi-D_J)
+F_{m5}(A^m-J^m) 
-\big[\la_2(\la_1-\la_J)+h.c. \big]
\nn\\
&& \hspace{-10pt} +\; (C_2-C_J)\big[
\p_5(X_3-\p_5\Phi)-\p_m\p^m\Phi \big]
\nn\\
&& \hspace{-20pt} +\; \Big[
\frac{i}{2}X_{12}^\ast(M_2-M_J)
-(\p_5\la_1+i\si^m\p_m\labar_2)(\chi_2-\chi_J)
+h.c.\Big] .
\eea
Without the WZ gauge imposed, the boundary action then
depends explicitly on the gauge degrees of freedom, $C_2$,
$\chi_2$, and $M_2$. The way to eliminate them without
imposing a gauge is to use a part of the boundary conditions
contained in $\bf V_2\beq J$,
\bea
C_2\beq C_J, \quad 
\chi_2\beq\chi_J, \quad
M_2\beq M_J .
\eea
This way we arrive at the action (\ref{SJY}).
Having used some of the boundary conditions in the action,
we expect that we would need to use boundary conditions in
checking supersymmetry of the simplified action.
As $\bf (V_2-J)$ is a gauge invariant vector superfield,
we have
\bea
\da_\eta {\bf (V_2-J)} = (\eta Q+\etabar\Qbar) {\bf (V_2-J)} ,
\eea
so that its components vary according to eq.~(\ref{bare}).
For example,
\bea
\da_\eta (C_2-C_J) &=& i\eta(\chi_2-\chi_J)+h.c. \nn\\[5pt]
\da_\eta (\chi_2-\chi_J) &=& 
\si^m\etabar\big[ \p_m (C_2-C_J)+i (A_m-J_m) \big]
+\eta (M_2-M_J) \nn\\[5pt]
\da_\eta (M_2-M_J) &=& 2\etabar\big[
(\labar_1-\labar_J)+i\sibar^m\p_m(\chi_2-\chi_J) \big] .
\eea
This implies that if we use the boundary conditions for
$C_2$ and $\chi_2$ in the action, then supersymmetry of the action 
requires using the $A_m$ boundary condition. Using the $M_2$
boundary condition leads to the $\la_1$ boundary condition.
And so on. This is indeed the pattern we observed in explicit
calculations. A final remark is that on-shell $X_{12}=0$, which
lets us avoid using the $M_2$ boundary condition. This is the
reason why supersymmetry of the bulk-plus-boundary action (\ref{SJY}) 
requires the use of only the $A_m$ boundary condition on-shell
(see eq.~(\ref{susyS})).

%%%%%%%%%%%%%%%%%%%%%%%%%%%%%%%%
%\newpage

%%%%%%%%%%%%%%%%%%%%%%%%%%%%%%%%

%%%%%%%%%%%%%%%%%%%%%%%%%%%%%%%%%%%%%%%%%%%%%%%%%%%%%%%%%%%%%%%
\end{document}